\documentstyle[twocolumn,aps,epsfig]{revtex}  
\begin{document}

\newcommand{\wt} {\widetilde}

\twocolumn[\hsize\textwidth\columnwidth\hsize\csname@twocolumnfalse\endcsname

\title{
Coulomb Suppression of NMR Coherence Peak in \\
Fullerene Superconductors
}
\author{Han-Yong Choi}
\address{Department of Physics,
Sung Kyun Kwan University, Suwon 440-746, Korea}

\date{Oct. 20, 1997}
\maketitle

\begin{abstract} 
The suppressed NMR coherence peak in the fullerene superconductors is
explained in terms of the dampings in the superconducting state
induced by the Coulomb interaction between conduction electrons. The
Coulomb interaction, modelled in terms of the onsite Hubbard
repulsion, is incorporated into the Eliashberg theory of
superconductivity with its frequency dependence considered 
self-consistently at all temperatures. The vertex correction is also
included via the method of Nambu. The frequency dependent
Coulomb interaction induces the
substantial dampings in the superconducting state
and, consequently, suppresses the anticipated NMR coherence peak of
fullerene superconductors as found experimentally.

\end{abstract}


]

It is generally accepted that the 
superconducting properties of the fullerene superconductors can be
understood in terms of phonon-mediated $s$-wave pairing
\cite{hebard,gunnarsson,ramirez,gelfand}. 
The $1/(T_1 T)$ for $s$-wave superconductors, where $T_1$ is the
nuclear spin-lattice relaxation time and $T$ is the temperature, is
expected to show a peak as $T$ is lowered below the transition
temperature, $T_c$, and is constant above $T_c$. It is referred to as
coherence peak or Hebel-Slichter peak. 
The expected coherence peak is
found substantially suppressed for fullerene superconductors
\cite{tycko,sasaki,pennington}. The underlying mechanism of
the suppression, however, is not clearly understood yet. 
The present Letter addresses the problem of NMR coherence peak
suppression in fullerene superconductors and explains it in terms of
the dampings in the superconducting state induced by the
frequency dependent screened Coulomb interaction between conduction
electrons.

The maximum of the normalized relaxation rate, $R_s/R_n$, 
in the limit of zero applied magnetic field 
was estimated to be $1.1 - 1.2$ 
for fullerene superconductors \cite{sasaki,pennington},
where $R = T_1^{-1}$ is the relaxation rate, and the subscripts $s$
and $n$, respectively, refer to the superconducting and normal states.
A similar behavior was also observed in $\mu$SR (muon spin relaxation)
experiments \cite{kiefl}. An analysis based on the
phonon-mediated Eliashberg theory \cite{akis} gives $T_c /\omega_{ph}
\approx 0.2$, and the characteristic phonon frequency $\omega_{ph} \approx
100~ cm^{-1}$ \cite{pennington}. This seems inconsistent with the
commonly held view of $\omega_{ph} \sim 1000~ cm^{-1}$ for fullerene
superconductors \cite{hebard,gunnarsson,ramirez,gelfand},
which implies $( R_s/R_n )_{max} \sim 2 - 3$ \cite{akis}. 
The increase of $(R_s/R_n)_{max}$ relative to
its normal state value, $0.1 - 0.2$, is therefore suppressed by an
order of magnitude from the expected value. Closely related is the
observation of the unexpected sensitivity of $R_s/R_n$ to the applied
magnetic field. The coherence peak is found completely
suppressed at only around 5 T for
Rb$_2$CsC$_{60}$, which is at least an order of
magnitude smaller than the expected value \cite{pennington}. It
may be understood on the ground that the zero field coherence peak is
substantially suppressed as mentioned above. A simple weighted average
of the suppressed intrinsic rate from the superconducting
region and the normal rate from the normal region inside the vortex
core will give the experimentally observed field dependence of $R_s
/R_n$ \cite{pennington}.

In order to understand the coherence peak suppression at zero magnetic
field, let us list the possible factors that are known to affect the NMR
coherence peak of $R_s/R_n $
\cite{nmrrev,choi1}, and check if any of those can explain the
suppression. The NMR coherence peak suppression may be attributed
to (a) momentum anisotropy of superconducting gap including non-$s$-wave 
pairing, (b) time reversal symmetry breaking such as magnetic
impurities or applied magnetic field, and/or (c) the damping effects
in superconducting state. For fullerene superconductors, the
superconductivity is of the phonon-mediated $s$-wave pairing, and due
to the orientational disorder of C$_{60}$ molecules, the fermi surface
anisotropy is not strong enough to suppress the coherence peak
\cite{choi1}. The time reversal symmetry breaking can not explain the
suppression either because there are no magnetic impurities in the
fullerene superconductors, and
we are considering the case of zero applied magnetic field. Because
either the gap anisotropy or time reversal symmetry breaking can not
explain the coherence peak suppression in the fullerene
superconductors, the near absence of the coherence peak
should be due to the $damping~ effects$.

The dampings of an
electron come from the scatterings of the electron with the phonons,
impurities, and/or other electrons. The damping from the scatterings
of electrons with the phonons, that is, the electron-phonon
interactions, is not strong enough: The
dimensionless electron-phonon coupling constant, $\lambda$, of the
fullerenes is estimated to be $0.5 - 1$ \cite{gunnarsson,ramirez,gelfand},
far smaller than $\sim 2$ needed to
suppress the NMR coherence peak \cite{allen}.
The far infrared reflectivity measurements of DeGiorge
$et~al.$ \cite{degiorge} show that the ratio $2 \Delta_0/k_B T_c
\approx 3.44-3.45$ both for K$_3$C$_{60}$ and Rb$_3$C$_{60}$. It is
very close to the BCS value of 3.52 and implies that
$\lambda$ can not be as large as 2. The impurity scatterings smear
out the fermi surface anisotropy and can not suppress the coherence
peak as explained previously. There remains, therefore, only one
possibility for inducing the dampings required to suppress the NMR
coherence peak in the fullerene superconductors: The scatterings
between electrons due to the $Coulomb~interactions$. This idea is
indeed verified in our detailed Eliashberg-Nambu (EN) calculations 
as will be detailed below.

The NMR relaxation rate for a superconductor with a finite bandwidth
of $B$ is given by
\begin{eqnarray}
\frac{1}{T_1 T} & \propto& \int_0^{\infty} d\epsilon \frac{\partial
f_F(\epsilon)}{\partial\epsilon}
\left\{ \left( Re \frac{\epsilon \theta(\epsilon)}
{\sqrt{\epsilon^2 -\Delta(\epsilon)^2}} \right)^2 \right.
\nonumber \\
&+& \left. \left( Re \frac{\Delta(\epsilon) \theta(\epsilon)}
{\sqrt{\epsilon^2 -\Delta(\epsilon)^2}} \right)^2 \right\}, 
\label{t1t}
\end{eqnarray}                                             
where $f_F (\epsilon) = 1/(1+e^{\beta\epsilon})$ is the
fermi distribution function, $\beta = 1/k_B T$, $ \theta(\epsilon)
= \tan^{-1} \left(
B /2 Z(\epsilon) \sqrt{\epsilon^2 -\Delta(\epsilon)^2}
\right)$, and $Z(\omega)$ and $\Delta(\omega)$ are, respectively, the
renormalization and gap functions. The finite conduction bandwidth
with a constant density of states (DOS) is explicitly considered
through the factor of $\theta$, which is $\pi/2$ for the usual case of
infinite bandwidth superconductors. For fullerene superconductors,
the fermi energy $\epsilon_F = B/2 \approx 0.2 - 0.3 $ eV and the
average phonon frequency $\omega_{ph} \approx 0.05 - 0.15$ eV. Consequently,
$\omega_{ph}/\epsilon_F \sim 1$ for fullerenes unlike conventional
metals where $\omega_{ph}/\epsilon_F \ll 1$. When
$\omega_{ph}/\epsilon_F \sim 1$, the phonon
vertex correction becomes important because the Migdal theorem
does not hold, and the Coulomb interaction should be considered more
carefully because the validity of the Coulomb pseudo-potential, $\mu^*
= \mu/[1+\mu \ln(\epsilon_F/\omega_{ph})]$, is unclear. In the
present work concerned with the narrow bandwidth superconductor of
fullerenes, therefore, the vertex correction is incorporated into the
Eliashberg theory via the method of Nambu \cite{nambu}. 
The Coulomb interaction is
modelled in terms of the onsite Hubbard repulsion and is included in
the theory with its frequency dependence considered 
self-consistently. As far as we are aware of, this is the first
self-consistent calculation of $1/(T_1 T)$ with the Eliashberg
formalism including the Coulomb
interaction and the vertex correction. The frequency dependence of
the screened Coulomb interaction, which comes from the polarization
diagrams of both the normal and pairing processes of renormalized
electrons, is important because the electron-phonon and Coulomb
interactions vary on a comparable frequency scale for a narrow
bandwidth superconductor of $\omega_{ph}/\epsilon_F \sim 1$.

The NMR coherence peak below $T_c$ is due to the increased DOS in the
superconducting state, that is, due to
the smallness of the denominator of Eq.\ (\ref{t1t}) when $\epsilon
\approx \Delta(\epsilon)$. In order for the peak to be suppressed,
therefore, the vanishingly small denominator should be avoided. The
$\sqrt{\epsilon^2 -\Delta(\epsilon)^2}$ of Eq.\ (\ref{t1t})  may not
vanish when there is a damping, that is, non-zero $\Delta_2
(\epsilon)$ for $\epsilon \approx \Delta_1 (\epsilon)$, where
$\Delta_1$ and $\Delta_2$ are, respectively, the real and imaginary
parts of the gap function.
The dampings in the superconducting state responsible for the NMR
coherence peak suppression is greatly increased when the frequency
dependence of the screened Coulomb interaction is retained
for $\omega_{ph}/\epsilon_F \sim 1$.
There have been several
works which emphasize the importance of including both the
electron-phonon and electron-electron interactions in understanding the
fullerenes \cite{knupfer}.

The EN equation can be written in the Matsubara frequency as:
\begin{eqnarray}
Z_n p_n &=& p_n +\frac{1}{\beta} \sum_m \left[ \lambda_{ph}(n-m)
-\lambda_{ch} (n-m) \right.
\nonumber \\
&+& \left. \lambda_{sp} (n-m) \right] 
\frac{2\theta_m Z_m p_m} {\sqrt{p_m^2 +\Delta_m^2}}  
+ \frac{1}{\pi \tau} \frac{\theta_n p_n} {\sqrt{p_n^2 +\Delta_n^2}},
\nonumber\\
Z_n \Delta_n &=& \frac{1}{\beta} \sum_m \left[ \lambda_{ph}(n-m)
-\lambda_{ch} (n-m) \right.
\nonumber \\
&-& \left. \lambda_{sp} (n-m) \right] 
\frac{ 2\theta_m Z_m \Delta_m }{\sqrt{p_m^2 +\Delta_m^2}} 
+ \frac{1}{\pi \tau} \frac{\theta_n \Delta_n} {\sqrt{p_n^2
+\Delta_n^2}}, 
\label{eli-imag}
\end{eqnarray}   
where $p_n = \pi T (2n+1)$ is the Matsubara frequency,
$\theta_n = \tan^{-1} \left( B/2 Z_n
\sqrt{p_n^2 +\Delta_n^2} \right)$, and $\lambda_{ph} (n-m)
=\int_0^{\infty} d\Omega \alpha^2 F(\Omega) 2 \Omega /
\left[\Omega^2 +(p_n -p_m )^2 \right]$ 
is the pairing kernel due to the electron-phonon interaction. 
Eq.\ (\ref{eli-imag}) is of the same form as the theory used to study
the spin fluctuation effects on superconductivity \cite{berk}. The
$\lambda_{ch} (n-m)$ and $\lambda_{sp} (n-m)$ are, respectively, the
interactions in the charge and spin channels 
due to the Hubbard repulsion, and are
determined self-consistently as
\begin{eqnarray}
\lambda_{ch} (k) &=& U N_F \left\{ 1/2- (\chi_n  +\chi_s ) \right.   
\nonumber \\
&+& \left. (\chi_n +\chi_s)^2 
\ln \left[1 +1/(\chi_n +\chi_s) \right] \right\} ,  
\nonumber \\
\lambda_{sp} (k) &=& U N_F \left\{ 1/2 +(\chi_n -\chi_s ) \right.
\nonumber \\
&+& \left. (\chi_n -\chi_s)^2 \ln[1 -1/(\chi_n -\chi_s)] \right\},
\label{lambda}
\end{eqnarray}
where $\chi_n (k)$ and $\chi_s (k)$ are the dimensionless
susceptibilities from, respectively, the normal and pairing 
processes given by
\begin{eqnarray}
\chi_n (k) &=& \frac{N_F U}{ \epsilon_F}
\frac{1}{\beta}\sum_l \theta_l \theta_{k+l} 
\frac{p_l p_{k+l}}{\sqrt{p_l^2 +\Delta_l^2} \sqrt{p_{k+l}^2
+\Delta_{k+l}^2} }, \nonumber \\
\chi_s (k) &=& \frac{N_F U}{ \epsilon_F}
\frac{1}{\beta}\sum_l \theta_l \theta_{k+l}
\frac{\Delta_l \Delta_{k+l}}
{\sqrt{p_l^2 +\Delta_l^2} \sqrt{p_{k+l}^2
+\Delta_{k+l}^2} }.
\label{chi}
\end{eqnarray}  
The $Z_m$ on the right hand side of Eq.\ (\ref{eli-imag}) 
represents the vertex correction of Nambu, which we
take as its form in the normal state for simplicity \cite{nambu}.
Including the vertex correction enhances the transition temperature in
accord with the previous works \cite{takada}. If we neglect the vertex
correction, the $Z_m$ should be put equal to 1. 
Self-consistent solution of Eq.\ (\ref{eli-imag}) together with
Eqs.\ (\ref{lambda}) and (\ref{chi}) gives 
$Z(ip_n)$ and $\Delta(ip_n)$ in the imaginary
frequency. To obtain $Z(\omega)$ and $\Delta(\omega)$ in the real
frequency, we perform the analytic continuations using the iterative
method of mixed representations \cite{choi1,marsiglio}. It is more
efficient than solving the Eliashberg equation directly in the real
frequency. The details of the Eliashberg-Nambu
formulation in the imaginary frequency and its analytic
continuation will be reported separately.

The EN equation of Eq.\ (\ref{eli-imag}) is solved
self-consistently via iterations with a set of the phonon spectral
function $\alpha^2 F(\Omega)$, Hubbard repulsion $U$, and impurity
scattering rate $\tau^{-1}$. To model the fullerene superconductors,
we take $\alpha^2 F(\Omega) = \sum_{\nu=1}^3 \alpha_{\nu}^2
F_{\nu}(\Omega)$,        
where $F_{\nu} (\Omega)$ is the truncated Lorentzian centered at 
$\omega_{\nu}$ with the broadening $\Gamma = \omega_{\nu}/5$, the
cutoff frequency $ \Gamma_c = 3 \Gamma$, and 
$\int_0^{\infty} d\Omega F_{\nu} (\Omega) = 1$ \cite{choi1}. Various
theoretical and experimental estimates
do not agree well with each other in terms of distribution of
coupling strength $\alpha_{\nu}^2$ among the different modes. 
These estimates show,
however, that the phonon spectra derived from the intramolecular $A_g$
and $H_g$ modes are distributed over $0.03 - 0.2$ eV with the total
$\lambda$ in the range $0.5 - 1$ \cite{gunnarsson,ramirez,gelfand}.
In view of this, we represent the phonon modes with three groups
centered around $ \omega_{\nu}$ = 0.04, 0.09, 0.18 eV, and
$ 2 N_F \alpha_{\nu}^2 /\omega_{\nu} = 0.3 ~\lambda_s,
0.2 ~\lambda_s, 0.5~ \lambda_s$, respectively, for $\nu$ = 1, 2, 3. 
Note that $\sum_{\nu=1}^3 2N_F \alpha_{\nu}^2 /\omega_{\nu} =
\lambda_s$. This choice of $\alpha^2 F(\Omega)$ gives the
logarithmically averaged phonon frequency $\omega_{ln} 
\approx 0.094$ eV,
which is a representative value of the various estimates of
$\omega_{ln}$. The $\lambda_s$ is set to give $T_c \approx 20 -
40$ K for a given $U$.
We take the fermi energy $\epsilon_F = B/2 = 0.25$ eV in the present
calculations, which gives the ratio $\omega_{ln} /\epsilon_F \approx
0.38$.

We put $\tau^{-1}$ = 0 for simplicity because the results are
insensitive to the impurity scatterings. 
Even though the Anderson theorem does not hold exactly
because of the finite bandwidth and the Hubbard repulsion in the
present theory, the thermodynamic properties are still insensitive
to the impurity scatterings.
We take $UN_F$ = 0.31 and $\lambda_s$ = 0.71 in the numerical
calculations reported below. We find $T_c$ = 0.0031
eV and $2 \Delta_0/k_B T_c$ = 3.1. It is interesting to note that
the present theory gives rather small $2 \Delta_0/k_B T_c$ value
which lies at the low end of the various estimates of the gap values. 
In the present
study, the long wavelength contribution of the Coulomb interaction is
not considered explicitly. Therefore, $U$ should be taken as a
screened value. The previous estimates give $UN_F \approx 0.3 - 0.4$
\cite{gunnarsson}.

Fig.\ 1 shows the superconducting gap function $\Delta(\omega)$ as a
function of $\omega$ at $T$ = 0.001 eV obtained by solving the
Eliashberg-Nambu equation. We took 220 Matsubara frequencies 
to solve Eq.\ (\ref{eli-imag}) by iterations and disregarded 20 high
frequency data to avoid boundary effects. The analytic continuations
were carried out with 2000 frequencies in the range between
$0 - 0.6$ eV. Fig.\ 1(a) is for an interacting system of $ UN_F$ =
0.31 and $\lambda_s$ = 0.71, and 1(b) is for a non-interacting case of
$U$ = 0 and $\lambda_s$ = 0.35, which gives $T_c$ = 0.0032 eV and
$2\Delta_0/k_B T_c = 3.7$. 
The solid and dashed lines, respectively,
represent the real and imaginary parts of $\Delta (\omega)$. 
The three peaks in $\Delta(\omega)$ reflect the three peaks in the
phonon spectral function $\alpha^2 F(\Omega)$. Note the
difference in $\Delta_2 (\omega)$ between the two cases for $\epsilon
\approx \Delta_1 (\epsilon)$: For the interacting case (a), the
superconducting state has quite a strong damping of $|
\Delta_2/\Delta_1| \approx 0.05$ around $\epsilon \approx \Delta_1
(\epsilon)$ due to the Coulomb interaction even at the low temperature
of $T/T_c \approx 0.3$, while $\Delta_2/\Delta_1 = 0$ around
$\epsilon \approx \Delta_1 (\epsilon)$ for the non-interacting case of
(b) because the thermal fluctuations are quenched at this
temperature. As $T$ is increased, $ |\Delta_2/\Delta_1 |$ is further
increased due to the thermal fluctuations. Substantial $\Delta_2$ is
what suppresses the NMR coherence peak as shown in Fig.\ 2.

\vspace{-0.1in}
\begin{figure}
\centerline{\epsfig{file=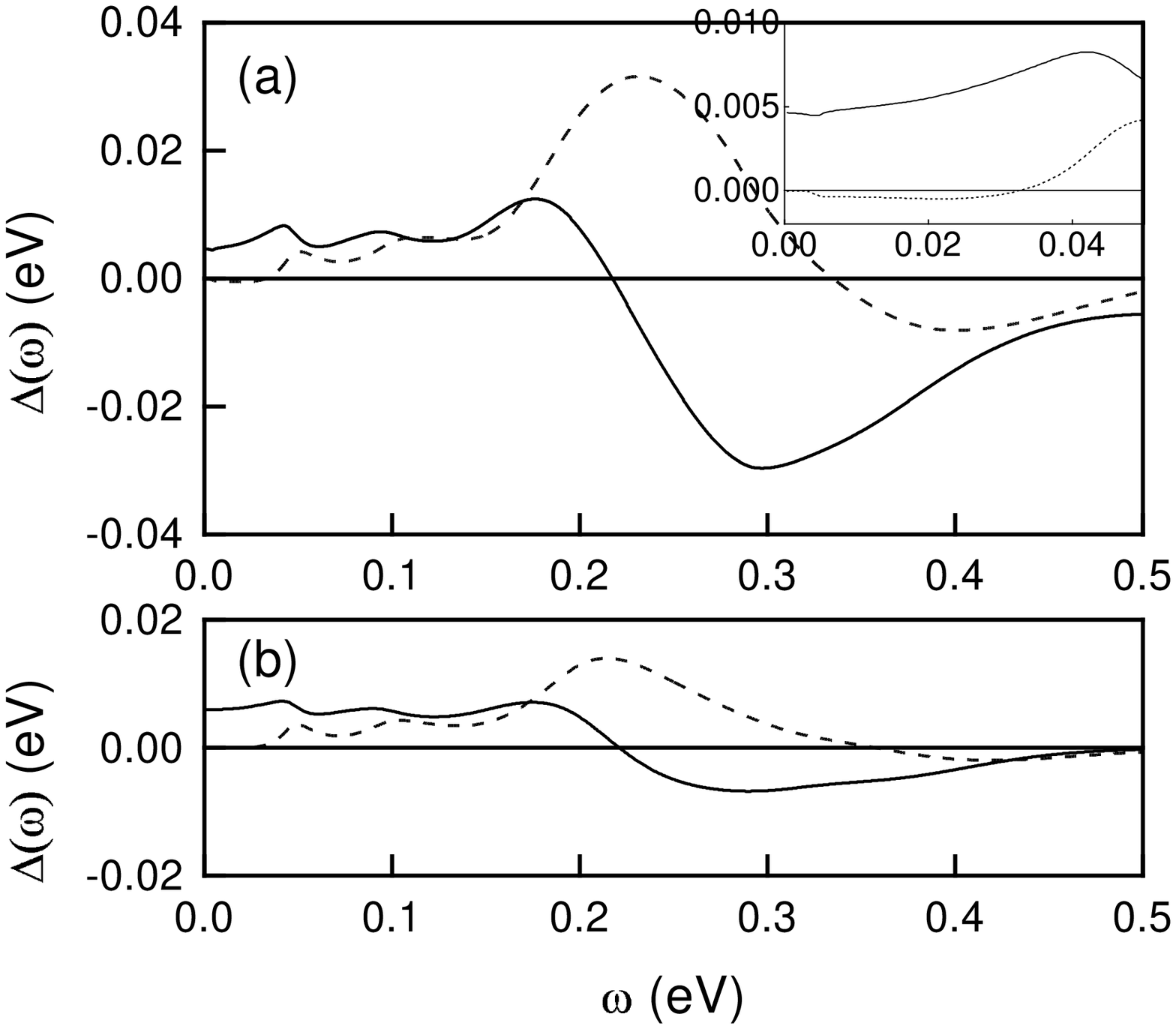,width=\linewidth}}
\vspace{-2in}
\caption{
The superconducting gap function, $\Delta(\omega)$, at $T= 0.001$ eV.
The solid and dished lines, respectively, stand for the real and
imagninary parts. We take $UN_F = 0.31$ and $\lambda_s = 0.71$ for
Fig.\ 1(a), and $U$ = 0 and $\lambda_s = 0.35$ for 1(b), so that the
two cases have similar $T_c$. The substantial damping of 
$|\Delta_2/\Delta_1 | \approx 0.05$ around $\epsilon \approx \Delta_1
(\epsilon)$ for the interacting case of (a) can be seen more clearly
in the inset of Fig.\ 1(a).
}
\label{fig1}
\end{figure}

The $\Delta(\omega)$ and $Z(\omega)$ obtained above
are then used to calculate the relaxation rate $T_1$ 
using Eq.\ (\ref{t1t}).
We show in Fig.\ 2 the normalized relaxation rate 
$R_s/R_n$ as a function of the reduced temperature $T/T_c$.   
The solid line is for $UN_F$ =
0.31 and $\lambda_s$ = 0.71 corresponding to Fig.\ 1(a). As expected,
the substantial damping in the superconducting state in the
interacting system suppresses the NMR coherence peak so that the
maximum of $R_s/R_n \approx 1.15$. By comparison, the corresponding
curve for the non-interacting case of
Fig.\ 1(b) ($U = 0$ and $\lambda_s = 0.35$) exhibits a much more
pronounced coherence peak as shown by the dotted curve, a hallmark of
the weak-coupling $s$-wave BCS superconductors.
It is clear that the strong Coulomb interaction can suppress the NMR
coherence peak for phonon-mediated $s$-wave superconductors with a
modest electron-phonon coupling constant. As $U$ is increased, the NMR
coherence peak is further suppressed. The dashed curve shows $R_s/R_n$
for $UN_F = 0.4$ and $\lambda_s = 0.8$, which gives $T_c \approx
0.003$ eV. We now show some of the available experimental data against
the theoretical curves in Fig.\ 2. The filled circles and squares
represent, respectively, the data by Stenger $et~al.$ from $^{13}$C
NMR of Rb$_2$CsC$_{60}$ at $B=1.5$ T and $B=3$ T, and the open up and
down triangles represent the data by Sasaki $et~al.$ from $^{13}$C NMR
of K$_3$C$_{60}$ at $B=2.93$ T. It seems that the experimental data
can be well described by $UN_F \approx 0.3 - 0.4$ and $\lambda_s
\approx 0.7 - 0.8$. We note that the Sasaki's data show somewhat more
suppressed NMR coherence peak compared with the Stenger's data, while
the Stenger's data show a bit narrow width compared with the
theoretical calculations.

\vspace{-0.1in}
\begin{figure}
\centerline{\epsfig{file=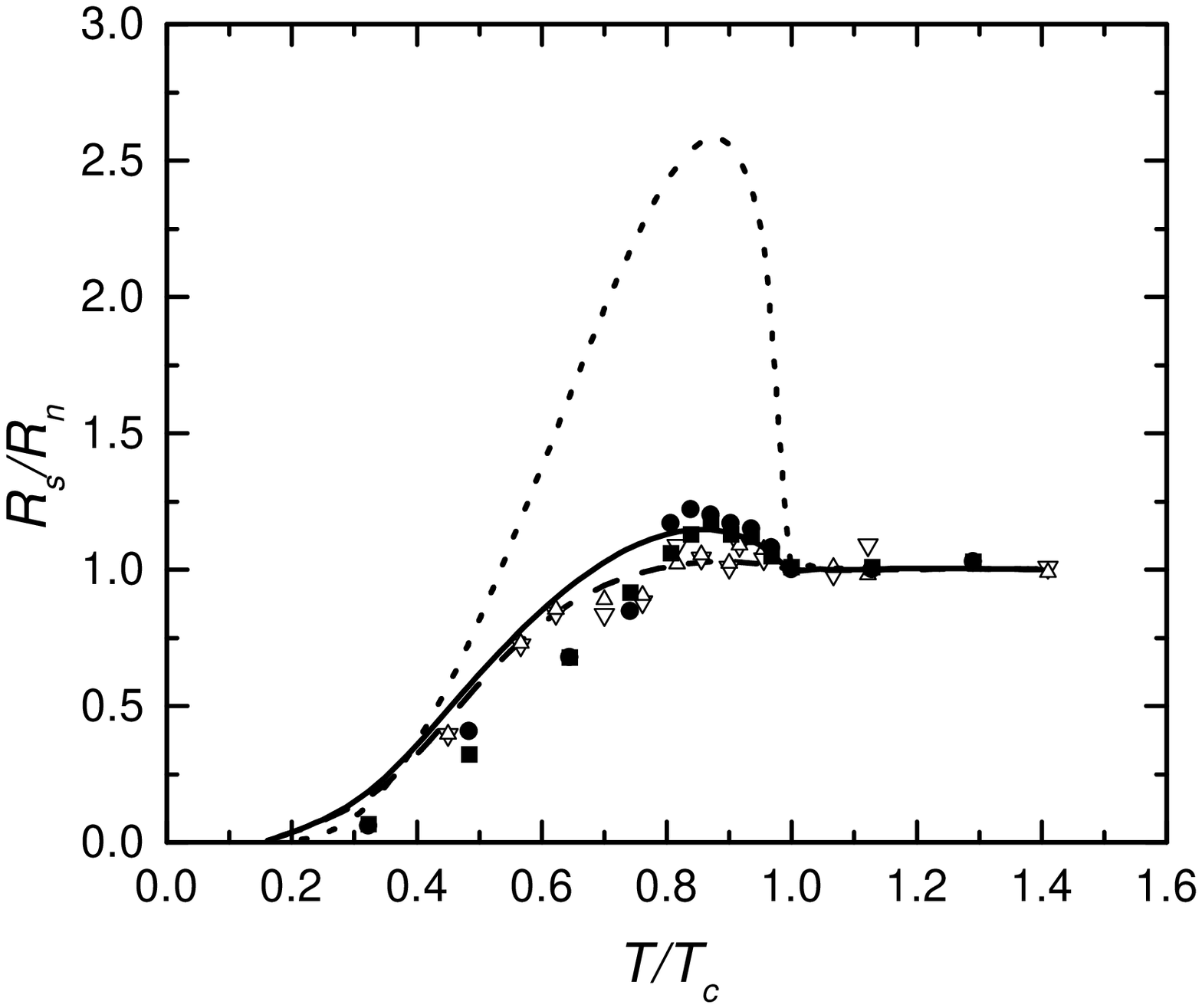,width=\linewidth}}
\vspace{-2.2in}
\caption{
The normalized relaxation rate, $R_s/R_n$, as a
function of the reduced temperature $T/T_c$. The solid curve is for
$UN_F = 0.31$ and $\lambda_s = 0.71$ corresponding to Fig.\ 1(a),
the dotted curve for non-interacting case of $UN_F=0$ and 
$\lambda_s=0.35$ corresponding to Fig.\ 1(b),
and the dashed curve is for $UN_F = 0.4$ and $\lambda_s = 0.8$.
The substantial dampings in the superconducting state for the
interacting cases suppress the NMR coherence peak as can clearly be
seen by comparing above curves. 
The filled circles and squares are the data from Stenger $et~al.$, and
the open up and down triangles are those from Sasaki $et~al.$
See the text for more detailed discussions.
}
\label{fig2}
\end{figure}

We point out that Mazin $et~al.$ have analyzed the NMR data using the
Eliashberg theory with two peaks in $\alpha^2 F(\Omega)$, one high
frequency peak at around $1000~ cm^{-1}$ of corresponding $\lambda
= 0.5$ and another very low frequency peak around $40~cm^{-1}$
of $\lambda = 2.7$ \cite{mazin}. Such a low frequency mode with
the strong coupling gives $\omega_{ln} = 66~cm^{-1}$ in sharp
contrast with the commonly held view of $\omega_{ln} \sim
1000~cm^{-1}$. The present work explains the NMR coherence peak
suppression without such a low frequency mode, and confirms that the
BCS-Eliashberg framework of $\omega_{ph} \sim 1000~cm^{-1}$ and
$\lambda \sim 0.5 -1$ can provide a consistent description of the wide
range of experimental data including $T_1$, provided that the
frequency dependent Coulomb interaction and the vertex correction are
properly incorporated.

To summarize: We have extended the standard Eliashberg theory for
narrow bandwidth superconductors by including the frequency dependent
screened Coulomb interaction together with the electron-phonon
interaction, and by including the vertex correction via the
Nambu's method. We then solved the Eliashberg-Nambu equation 
self-consistently at all temperatures to obtain the gap and
renormalization functions, $\Delta(\omega)$ and $Z(\omega)$,
respectively, which are used to calculate the nuclear spin-lattice
relaxation rate, $T_1^{-1}$. The frequency dependent Coulomb
interactions between conduction electrons induce the substantial
dampings in the superconducting state and, consequently, suppress the
NMR coherence peak in the fullerene superconductors. 
The present work, therefore, has shown that the $T_1$ experiments can
be understood with the view of 
$\omega_{ph} \sim 1000~cm^{-1}$ and $\lambda \sim 0.5 - 1$.
It remains to be seen if other experimental data can also be
understood with the view.

The author would like to thank Drs.\ Charles Pennington and 
Susumu Sasaki for useful discussions, 
and acknowledge the support by Korea Science and
Engineering Foundation (KOSEF) through Grant No.\ 96-0702-02-01-3 and
through the Center for Theoretical Physics, Seoul National University,
and by the Ministry of Education through Grant No.\ BSRI-97-2428.


\begin{references}  

\bibitem{hebard} A. F. Hebard, Phys. Today {\bf 45} (11), 26 (1992).
\bibitem{gunnarsson} O. Gunnarsson, Rev. Mod. Phys. {\bf 69}, 575
(1997). 
\bibitem{ramirez} A. P. Ramirez, Superconductivity Rev. 
{\bf 1}, 1 (1994).
\bibitem{gelfand} M. P. Gelfand, Superconductivity Rev. {\bf 1}, 103
(1994).
\bibitem{tycko} R. Tycko {\it et al.}, Phys. Rev. Lett. {\bf 68}, 1912
(1992). 
\bibitem{sasaki} S. Sasaki, A. Matsuda, and C. W. Chu, J. Phys. Soc.
Jpn. {\bf 63}, 1670 (1994).
\bibitem{pennington} V. A. Stenger {\it et al.}, Phys. Rev. Lett. {\bf
74}, 1649 (1995); C. H. Pennington and V. A. Stenger, Rev. Mod.
Phys. {\bf 68}, 855 (1996), and references therein. 
\bibitem{kiefl} R. F. Kiefl {\it et al.}, Phys. Rev. Lett. {\bf 70}, 
3987 (1993); W. A. MacFarlane {\it et al.}, submitted to Phys. Rev. B
(1998).
\bibitem{akis} R. Akis, C. Jiang, and J. P. Carbotte, Physica C {\bf 176},
485 (1991). 
\bibitem{nmrrev} D. E. McLaughlin, in {\it
Solid State Physics: Advances in Research and Applications}, ed. H.
Ehrenreich {\it et al.} (Academic, New York, 1976) vol. 31, p. 1, and
references therein.     
\bibitem{choi1} H. Y. Choi, Phys. Rev. B {\bf
53}, 8591 (1996), and references therein.  
\bibitem{allen} P. B. Allen and D. Rainer, Nature {\bf 349}, 396 (1991);
Y. O. Nakamura {\it et al.}, Solid State Commun.
{\bf 78}, 393 (1991). 
\bibitem{degiorge} L. DeGiorge
{\it et al.}, Nature (London) {\bf 369}, 541 (1994); Phys. Rev. B
{\bf 49}, 7012 (1994). 
\bibitem{nambu} Y. Nambu, Phys. Rev. {\bf 117}, 648 (1960).      
\bibitem{knupfer} M. Knupfer and J. Fink, Phys. Rev. Lett. {\bf 79},
2714 (1997). 
\bibitem{berk} N. F. Berk and J. R. Schrieffer, Phys. Rev. Lett.
{\bf 17}, 433 (1966).
\bibitem{takada} Y. Takada, J. Phys. Chem.
Solids {\bf 54}, 1779 (1993). 
\bibitem{marsiglio} F. Marsiglio, M. Schossmann, and J. P.
Carbotte, Phys. Rev. B {\bf 37}, 4965 (1988).       
\bibitem{mazin} I. I. Mazin {\it et al.}, Phys. Rev. B {\bf 47},
538 (1993). 

                      
\end{references}
\end{document}